\RequirePackage{fix-cm}
\documentclass[reqno,draft]{amsproc}
    \numberwithin{equation}{section}
\usepackage{fixltx2e}
\usepackage[citation-order,nobysame]{amsrefs}
\usepackage{mathtools}
    \mathtoolsset{showonlyrefs}

\newcommand\preprint[1]{\begin{flushright} #1 \end{flushright}
    \vspace{2.25cm}}

\DeclareMathOperator*{\Asym}{\textrm{Asym}}

\renewcommand{\Re}{\mathrm{Re}\,{}}

\newcommand{\rme}{\mathrm{e}}
\newcommand{\rmi}{\mathrm{i}}
\newcommand{\rmd}{\mathrm{d}}
\newcommand{\eps}{\varepsilon}
\newcommand{\wt}{\widetilde}
\newcommand{\ot}{\mathop{\otimes}}
\newcommand{\caM}{\mathcal{M}}
\newcommand{\caN}{\mathcal{N}}
\newcommand{\caA}{\mathcal{A}}
\newcommand{\caE}{\mathcal{E}}
\newcommand{\bra}[1]{\langle #1 \rvert}
\newcommand{\ket}[1]{\lvert #1 \rangle}

\begin{document}

\preprint{DDF 441/12/07}

\title[Emptiness formation probability]
{Emptiness formation probability in the domain-wall
six-vertex model}

\author{F. Colomo}
\address{INFN, Sezione di Firenze\\
Via G. Sansone 1, 50019 Sesto Fiorentino (FI), Italy}
\email{colomo@fi.infn.it}

\author{A.G. Pronko}
\address{Saint Petersburg Department of V.A.~Steklov Mathematical
Institute of Russian Academy of Sciences\\
Fontanka 27, 191023 Saint Petersburg, Russia}
\email{agp@pdmi.ras.ru}

\begin{abstract}

The emptiness formation probability in the six-vertex model with domain
wall boundary conditions is considered. This correlation function allows
one to address the problem of limit shapes in the model. We apply the
quantum inverse scattering method to calculate the emptiness formation
probability for the inhomogeneous model. For the homogeneous model, the
result is given both in terms of certain determinant and as a multiple
integral representation.
\end{abstract}

\maketitle

\section{Introduction}

A special kind of fixed boundary conditions,
the so-called domain wall boundary conditions, was first considered
for the six-vertex model by Korepin in seminal paper \cite{K-82}.
In \cite{I-87}, Izergin showed that
the partition function of the model on the finite lattice
can be found exactly in terms of certain determinant; see also paper
\cite{ICK-92} for details. It was later shown in \cites{KZj-00,Zj-00},
by studying the thermodynamic limit,
that the free energy per site is different with respect to the case
of periodic boundary conditions. This fact hints at spatial
separation of phases (e.g., ferroelectric order and disorder), which
is confirmed both numerically \cites{E-99,SZ-04,AR-05} and analytically
\cites{Zj-02,PR-07}.

To get some details about the phase separation phenomena, e.g., to find
the shape of the spatial curve separating the phases, or limit shape,
one has to know some appropriate correlation function.
The problem of computing
the correlation functions in the `domain-wall' six-vertex model
has been addressed in papers \cites{BKZ-02,BPZ-02,FP-04,CP-05c} where some
correlation functions near the boundary were found.

In this paper, we continue the study of correlation functions of the six-vertex
model with domain wall boundary conditions. Specifically, we consider here
a particular non-local correlation
function, the emptiness formation probability (EFP). This function describes
the probability of having a set of consecutive
horizontal edges along a given column, all in a given state;
we consider here the case when the set starts from the top boundary
and extends inside the lattice.
This correlation function allows one to address the problem of
limit shapes in the model \cite{CP-07}.

To compute EFP, we follow the lines of papers \cites{BPZ-02,CP-05c,CP-05b} where
the quantum inverse scattering method (QISM) \cites{TF-79,KBI-93}
and some facts from the theory of orthogonal polynomials were used.
Mostly following ideas (as well as notations)
of \cite{CP-05c}, we represent EFP in certain determinantal form, which
is shown here to be also equivalent to some multiple integral.
This last representation recalls analogous multiple integral representations
for  correlation functions of quantum spin chains
\cites{JM-95,KMT-00,BKS-03,BJMST-06}.

The paper is organized as follows. In the next Section we start with
giving some definitions and fixing some notations. The quantum
inverse scattering method
in application to the model is considered in Section 3. The core calculation
of EFP for the inhomogeneous model is contained in Section 4.
The homogenous limit is performed in Section 5,
where the main result is given given both in terms of certain determinant and
as a multiple integral representation. Section 6 is devoted to
discussion of equivalent multiple integral representations for EFP.

\section{Some definitions and notations}

\subsection{The model}

The six-vertex model is a statistical
mechanics model in which the local states are associated with edges of a
square lattice, and the Boltzmann weights are assigned to its vertices.
The states can take two values, which are often denoted
by arrows pointing along the edge. Among the sixteen possible arrow configurations
around a vertex only six are allowed (having nonzero Boltzmann weights),
with equal number of incoming and outgoing arrows. In this paper we consider
the model on a lattice having both $N$ rows
and $N$ columns (`the $N\times N$ lattice') with the boundary states fixed in
a special way: all arrows on the left and right boundaries
are outgoing while on the top and bottom boundaries
all arrows are incoming. Such a model is
called the six-vertex model with domain wall boundary
conditions.

In the six-vertex model with invariance under  reversal of all
arrows there are three possible values for Boltzmann weights at each vertex,
usually denoted as $a$, $b$, and $c$. To use the quantum inverse
scattering method (QISM) for calculations we
will consider the inhomogeneous version of the model, in which
the weights of the vertex being
at the intersection of $k$-th horizontal line
and $\alpha$-th vertical line are
\begin{equation}\label{abc-im}
a_{\alpha k}=a(\lambda_\alpha,\nu_k),\qquad
b_{\alpha k}=b(\lambda_\alpha,\nu_k),\qquad
c_{\alpha k}=c,
\end{equation}
where
\begin{equation}\label{abc}
a(\lambda,\nu)=\sin(\lambda-\nu+\eta),\qquad
b(\lambda,\nu)=\sin(\lambda-\nu-\eta),\qquad
c=\sin(2\eta)
\end{equation}
and we enumerate vertical lines (labelled by Greek indices)
from right to left, and horizontal lines (labelled by Latin indices)
from top to bottom.
The parameters $\lambda_1,\dots,\lambda_N$ are assumed to be all different;
the same is assumed about $\nu$'s.
After applying QISM we set these parameters
equal within each set, $\lambda_\alpha=\lambda$ and
$\nu_k=\nu$ and, without losing generality, we can assume that $\nu=0$.
In this way we obtain the homogenous model; the indicated procedure will be
referred to as homogeneous limit.

The partition function of the inhomogeneous model is defined in a standard
way as the sum over all possible configurations, each configuration being
assigned its Boltzmann weight, whcih is the product of all vertex weights
over  the lattice,
\begin{equation}
Z_N= \sum_{\mathcal{C}}^{}
\prod_{\alpha=1}^{N}\prod_{k=1}^{N}w_{\alpha k}(C).
\end{equation}
Here $w_{\alpha k}(C)$ takes values $w_{\alpha k}(C)=
a_{\alpha k},b_{\alpha k},c_{\alpha k}$, depending on the configuration
$C$. Because of \eqref{abc-im},
$Z_N=Z_N(\lambda_1,\dots,\lambda_N;\nu_1,\dots,\nu_N)$ where
$\lambda$'s and $\nu$'s are regarded as `variables';
$\eta$ is regarded as a parameter (having the meaning of a `coupling constant')
and it is often omitted
in notations. In QISM the dependence on $\lambda$'s and
$\nu$'s play an important role (in particular, $Z_N$ is invariant under
permutations within each set of variables).

\subsection{QISM formulation}

We now define the main objects of QISM in relation to the model.
First, let us consider vector space $\mathbb{C}^2$ and denote its basis
vectors as the spin-up and spin-down states
\begin{equation}
\ket{\uparrow}=
\begin{pmatrix}
1 \\ 0
\end{pmatrix},\qquad
\ket{\downarrow}=
\begin{pmatrix}
0 \\ 1
\end{pmatrix}.
\end{equation}
To each lattice row and column we associate vector space $\mathbb{C}^2$.
We also use the convention that upward and right arrows correspond to
the `spin up'  state while downward and left arrows correspond
to the  `spin down' state.

Next, let us introduce the quantum $L$-operator,
which can be defined as a matrix of the Boltzmann weights.
Namely, to each vertex being intersection of the
$\alpha$-th vertical line (column) and the $k$-th horizontal line
(row) we associate the
operator $L_{\alpha k}(\lambda_\alpha,\nu_k)$ which acts in the direct
product of two vector spaces $\mathbb{C}^2$: in the `horizontal' space
$\mathcal{H}_k=\mathbb{C}^2$ (associated with the $k$-th row) and in the
`vertical' space $\mathcal{V}_\alpha=\mathbb{C}^2$ (associated with the
$\alpha$-th column). We regard
arrow states on the top and right edges of the
vertex as `in' indices of the $L$-operator while those on
the bottom and left edges as `out' ones. Explicitly, the $L$-operator
reads
\begin{equation}\label{Lop}
L_{\alpha k}(\lambda_\alpha,\nu_k)=
\sin(\lambda_\alpha-\nu_k+\eta\,\tau_\alpha^z \sigma_k^z )
+ \sin(2\eta)(\tau_\alpha^{-}\sigma_k^{+}
+\tau_\alpha^{+}\sigma_k^{-}).
\end{equation}
Here $\tau$'s ($\sigma$'s) are Pauli matrices of the corresponding vertical
(horizontal) vector spaces.

Further, we introduce the monodromy matrix, which is an
ordered product of $L$-operators.
We define monodromy matrix here as a product of $L$-operators
along a column, regarding
the corresponding vertical space $\mathcal{V}_\alpha$ as
an `auxiliary' space. The horizontal spaces
$\mathcal{H}_k$ will be regarded as `quantum' spaces;
the space $\mathcal{H}=\ot_{k=1}^N\mathcal{H}_k$ is therefore
the total quantum space. In defining the monodromy matrix
it is convenient to think of $L$-operator
as acting in $\mathcal{V}_\alpha\otimes\mathcal{H}$ and, moreover,
writing it as $2$-by-$2$ matrix in $\mathcal{V}_\alpha$,
with the entries being quantum operators (acting in $\mathcal{H}$),
\begin{equation}\label{L-op}
L_{\alpha k}(\lambda_\alpha,\nu_k)=
\begin{pmatrix}
\sin(\lambda_\alpha-\nu_k+\eta\,\sigma_k^z) &  \sin(2\eta)\,\sigma_k^-\\
\sin(2\eta)\,\sigma_k^+ & \sin(\lambda_\alpha-\nu_k-\eta\,\sigma_k^z)
\end{pmatrix}_{[\mathcal{V}_\alpha]}.
\end{equation}
Here the subscript indicates that this is a matrix in $\mathcal{V}_\alpha$ and
$\sigma_k^{l}$ ($l=+,-,z$) denote quantum operators
in $\mathcal{H}$ acting
as Pauli matrices in $\mathcal{H}_k$ and identically elsewhere.
The monodromy matrix is defined as
\begin{align}
T_\alpha(\lambda_\alpha)
& =
L_{\alpha N}(\lambda_\alpha,\nu_N) \cdots
L_{\alpha 2}(\lambda_\alpha,\nu_2) L_{\alpha 1}(\lambda_\alpha,\nu_1)
\notag\\ &
=\begin{pmatrix}
A(\lambda_\alpha)& B(\lambda_\alpha) \\
C(\lambda_\alpha)& D(\lambda_\alpha)
\end{pmatrix}_{[\mathcal{V}_\alpha]}.
\end{align}
The operators $A(\lambda)=A(\lambda;\nu_1,\dots,\nu_N)$, etc,
act in $\mathcal{H}$. They  play an important role in QISM.

Operators $A(\lambda)$, $B(\lambda)$, $C(\lambda)$, and $D(\lambda)$,
admit simple graphical
interpretation as columns of the lattice, with top and bottom
arrows fixed.
Let us introduce `all spins down' and `all spins up' states
\begin{equation}\label{UpDown}
\ket{\Uparrow}=\ot_{k=1}^{N}\ \ket{\uparrow}_k,\qquad
\ket{\Downarrow}=\ot_{k=1}^N\ \ket{\downarrow}_k,
\end{equation}
where $\ket{\uparrow}_k$ and $\ket{\downarrow}_k$
are basis vectors of $\mathcal{H}_k$. In the case of domain wall boundary
conditions each column corresponds to an operator $B(\lambda_\alpha)$
(where $\alpha$ is the number of the column) while
vectors \eqref{UpDown} describe states on the right and left
boundaries;
the partition function reads:
\begin{equation}\label{ZBBB}
Z_N= \bra{\Downarrow} B(\lambda_N) \cdots B(\lambda_2)B(\lambda_1)\ket{\Uparrow}.
\end{equation}

\subsection{Izergin-Korepin formula}

In \cite{K-82} Korepin established recursion relations for the partition function,
and in \cite{I-87} Izergin showed that these relations are
satisfied by the following explicit expression (see also \cite{ICK-92} for
details)
\begin{equation}\label{ZN}
Z_N=
\frac{\prod_{\alpha=1}^N \prod_{k=1}^N
a(\lambda_\alpha,\nu_k)b(\lambda_\alpha,\nu_k)}{
\prod_{1\leq\alpha<\beta\leq N}d(\lambda_\beta,\lambda_\alpha)
\prod_{1\leq j<k\leq N}d(\nu_j,\nu_k)}\,
\det\caM,
\end{equation}
where $\caM$ is $N$-by-$N$ matrix with entries
\begin{equation}\label{matT}
\caM_{\alpha k}=\varphi(\lambda_\alpha,\nu_k),\qquad
\varphi(\lambda,\nu)=\frac{c}{a(\lambda,\nu) b(\lambda,\nu)},
\end{equation}
while $a(\lambda,\nu)$, $b(\lambda,\nu)$ and $c$ are defined
in \eqref{abc}, and function $d(\lambda,\lambda')$, standing in the
pre-factor of \eqref{ZN}, is
\begin{equation}\label{d}
d(\lambda,\lambda'):=\sin(\lambda-\lambda').
\end{equation}
In the next Section we sketch a proof of \eqref{ZN}, originally
given in \cite{BPZ-02},
which uses exclusively the Yang-Baxter algebra.
The method is useful since it can be generalized
to the case of correlation functions (in contrast to the original approach of
Korepin and Izergin, \cite{ICK-92}).

In the homogenous limit, i.e., when $\lambda_\alpha=\lambda$ and
$\nu_k=0$, expression \eqref{ZN}
simplifies to
\begin{equation}\label{ZNhom}
Z_N=\frac{[\sin(\lambda-\eta)\sin(\lambda+\eta)]^{N^2}}
{\prod_{n=1}^{N-1}(n!)^2}\, \det\caN
\end{equation}
where $N$-by-$N$ matrix $\caN$ has entries
\begin{equation}\label{varphi}
\caN_{\alpha k}=\partial_{\lambda}^{\alpha+k-2}
\varphi(\lambda),\qquad
\varphi(\lambda):=\varphi(\lambda,0)=
\frac{\sin(2\eta)}{\sin(\lambda-\eta)\sin(\lambda+\eta)}.
\end{equation}
Expression \eqref{ZNhom} was given for the first time in \cite{I-87};
the derivation of  \eqref{ZNhom} from \eqref{ZN}
was explained in detail in \cite{ICK-92}.

\subsection{Emptiness formation probability}

Let us denote by $F_N^{(r,s)}$ the probability of having all arrows on $s$
first horizontal edges (counted, as usual, from the top of the lattice)
between $r$-th and $(r+1)$-th columns, to be all pointing left.
Using operator formalism we can define this probability as
\begin{equation}\label{EFP}
F_N^{(r,s)}
=Z^{-1} \bra{\Downarrow} B(\lambda_N) \cdots B(\lambda_{r+1})\,
\pi_1\cdots\pi_s\, B(\lambda_r)\cdots B(\lambda_1)\ket{\Uparrow}.
\end{equation}
Here $\pi_j$ denotes the projector on the spin-down state (which
correspondingly fixes  the arrow to be pointing left),
\begin{equation}
\pi_j=\tfrac{1}{2}(1-\sigma_j^z).
\end{equation}
We shall call correlation function \eqref{EFP} as emptiness formation
probability (EFP), adopting the name of  similar object
from the quantum spin chain context \cite{KBI-93}.

In this respect we comment that the name  `emptiness formation probability'
for the quantity defined by formula \eqref{EFP}
has to understood with some care, since
it is in fact corresponds to a different object.
Indeed,  if one regards, following the common practice in QISM, the vector
$\ket{\Uparrow}$ as an `empty' state and $B$'s as `creation' operators
over this state, then our definition actually corresponds rather
to some `fullness formation probability'. Nevertheless, we shall follow
the tradition and call this quantity EFP.

One can consider, instead of EFP defined by \eqref{EFP},
the true emptiness formation probability, defining it
by replacing all $\pi$'s in \eqref{EFP} by
$\bar\pi_j=1-\pi_j$. However, contrarily to the standard situation
in quantum spin chains in absence of external field, such quantity cannot
be related to EFP defined by \eqref{EFP}.
This is due to the fact that in the domain-wall six-vertex model
the local polarization is non-vanishing almost everywhere
over the lattice, or, in other words, the spin-reversal symmetry
is broken by the boundary conditions.

Our choice of correlation function \eqref{EFP} is motivated by
its further application to study limit shapes of the model.
Indeed, because of peculiarity of both
the domain-wall boundary conditions and the  six-vertex model
rule of conservation of  incoming and outgoing arrows through each
lattice vertex,
EFP \eqref{EFP} actually measures the probability
that all vertices in the top-left $(N-r)\times s$ sublattice
have the same configuration of arrows, namely, all arrows point to the
left or downwards. Hence EFP measures ferroelectric order or `freezing'
of states. The limit shape arises in some appropriate scaling
limit and corresponds to some  curve where EFP
jumps from one to zero as the size of this $(N-r)\times s$ sublattice
increases (see \cite{CP-07} for further details).

\section{QISM and recurrence relations}

\subsection{Yang-Baxter algebra}

One of the most basic relations of QISM is the so-called ``RLL'' relation
\cites{B-82,G-83,KBI-93}, which reads
\begin{equation}
R_{\alpha\alpha'}(\lambda,\lambda')
\big[L_{\alpha k}(\lambda,\nu)\otimes L_{\alpha' k}(\lambda',\nu)\big]=
\big[L_{\alpha k}(\lambda',\nu)\otimes L_{\alpha' k}(\lambda,\nu)\big]
R_{\alpha\alpha'}(\lambda,\lambda').
\end{equation}
Here $R_{\alpha\alpha'}(\lambda,\lambda')$, called the $R$-matrix,
is a matrix acting in the direct product of two auxiliary vector spaces,
$\mathcal{V}_\alpha\ot\mathcal{V}_{\alpha'}$, and it can be conveniently
represented as a $4$-by-$4$ matrix (we assume that the first space refers to the
$2$-by-$2$ blocks, while the second one to the entries in the blocks):
\begin{equation}
R_{\alpha\alpha'}(\lambda,\lambda')=
\begin{pmatrix}
f(\lambda',\lambda) & 0 &0 &0 \\
0 & g(\lambda',\lambda) &1 &0 \\
0 &1 & g(\lambda',\lambda) &0 \\
0 &0 &0 & f(\lambda',\lambda)
\end{pmatrix}_{[\mathcal{V}_\alpha\ot\mathcal{V}_{\alpha'}]}.
\end{equation}
Here the functions $f(\lambda',\lambda)$ and $g(\lambda',\lambda)$ are
\begin{equation}
f(\lambda',\lambda)=
\frac{\sin(\lambda-\lambda'+2\eta)}{\sin(\lambda-\lambda')},\qquad
g(\lambda',\lambda)=
\frac{\sin(2\eta)}{\sin(\lambda-\lambda')}.
\end{equation}
This $R$-matrix is also sometimes referred to as XXZ chain
$R$-matrix, due to relation of the six-vertex model
with Heisenberg XXZ quantum spin chain. It is to be mentioned
that here and below
we are mainly following notations and conventions of book \cite{KBI-93}.

The importance of the RLL relation above is that it implies
the following relation,
which, in turn, can be called RTT relation,
\begin{equation}\label{RTT}
R_{\alpha\alpha'}(\lambda,\lambda')
\big[T_\alpha(\lambda)\otimes T_{\alpha'}(\lambda')\big]=
\big[T_\alpha(\lambda')\otimes T_{\alpha'}(\lambda)\big]
R_{\alpha\alpha'}(\lambda,\lambda').
\end{equation}
This relation contains all commutation relations
between the operators $A(\lambda)$, $B(\lambda)$, $C(\lambda)$, and
$D(\lambda)$. The algebra of these operators is called
the Yang-Baxter algebra, or the quantum algebra of monodromy matrix.
Among the sixteen relations contained in \eqref{RTT}
the following two of them will be used explicitly below, namely,
\begin{equation} \label{BB}
B(\lambda)\, B(\lambda')=
B(\lambda')\, B(\lambda).
\end{equation}
and
\begin{equation} \label{AB}
A(\lambda)\, B(\lambda')=
f(\lambda,\lambda')\, B(\lambda')\, A(\lambda)
+g(\lambda',\lambda)\, B(\lambda)\, A(\lambda').
\end{equation}

\subsection{The `two-site model'}

Let us consider the following decomposition of the monodromy matrix
\begin{equation}
T(\lambda)=T_{2}(\lambda) T_{1}(\lambda),
\end{equation}
where $T_1(\lambda)$ is defined as a product of the first several
$L$-operators while $T_2(\lambda)$ is the product of the remaining
ones. Such a decomposition is sometimes called `two-site model' \cite{KBI-93}.
We shall consider here the case when $T_1(\lambda)$ consists of just
one $L$-operator,
\begin{equation}\label{T2T1}
T_2(\lambda)=
L_{\alpha N}(\lambda,\nu_N) \cdots
L_{\alpha 2}(\lambda,\nu_2),\qquad
T_1(\lambda)= L_{\alpha 1}(\lambda,\nu_1).
\end{equation}
Introducing the operators $A_1(\lambda)$, $A_2(\lambda)$, etc, as
operator-valued entries of the corresponding monodromy matrices
$T_1(\lambda)$, $T_2(\lambda)$, respectively, taking into account
that $B(\lambda)=A_2(\lambda)B_1(\lambda)+B_2(\lambda)D_1(\lambda)$, and
using \eqref{L-op}, we have
\begin{align}
B(\lambda)&= A_2(\lambda)\, c \sigma_1^- +
B_2(\lambda)\left[a(\lambda,\nu_1)\pi_1 + b(\lambda,\nu_1) \bar \pi_1\right]
\notag\\ & =
\begin{pmatrix}
b(\lambda,\nu_1)B_2(\lambda)& 0\\
c A_2(\lambda) & a(\lambda,\nu_1)B_2(\lambda)
\end{pmatrix}_{[\mathcal{H}_1]}.
\end{align}

The lower-triangle structure of operator $B(\lambda)$ as a matrix
in $\mathcal{H}_1$ leads to the property that the product of several
$B$'s has the form
\begin{equation}\label{EEE}
B(\lambda_n) \cdots B(\lambda_1) =
\begin{pmatrix}
E_{11}(\lambda_1,\dots,\lambda_n)&0\\
E_{21}(\lambda_1,\dots,\lambda_n)& E_{22}(\lambda_1,\dots,\lambda_n)
\end{pmatrix}_{[\mathcal{H}_1]}.
\end{equation}
The diagonal entries $E_{11}(\lambda_1,\dots,\lambda_n)$,
$E_{22}(\lambda_1,\dots,\lambda_r)$
are simply proportional to $B_2(\lambda_n)\cdots B_2(\lambda_1)$,
while the non-diagonal entry, $E_{21}(\lambda_1,\dots,\lambda_n)$, reads
\begin{multline}\label{E21}
E_{21}(\lambda_1,\dots,\lambda_n)= \sum_{\alpha=1}^{n}
\prod_{\beta=\alpha+1}^{n} a(\lambda_\beta,\nu_1)\cdot
c\cdot
\prod_{\beta=1}^{\alpha-1} b(\lambda_\beta,\nu_1)
\\ \times
B_2(\lambda_n)\cdots B_2(\lambda_{\alpha+1}) A_2(\lambda_\alpha)
B_2(\lambda_{\alpha-1}) \cdots B_2(\lambda_1).
\end{multline}
It is to be stressed that due to \eqref{BB}
entries in \eqref{EEE} are totally symmetric under permutations of $\lambda$'s.
While this is completely evident for the diagonal entries, such a property
is rather non-trivial for expression  \eqref{E21},
and is a consequence of  commutation relation \eqref{AB} for operators
$B_2(\lambda)$ and $A_2(\lambda)$.

\subsection{The key relation}

In dealing with the `two-site model' it is useful to consider the corresponding
decomposition of the vectors `all spins up' and `all spins down', e.g.,
$\ket{\Uparrow}=\ket{\Uparrow_1}\ot \ket{\Uparrow_2}$. To fit \eqref{T2T1}
we set $\ket{\Uparrow_1}=\ket{\uparrow}_1$ and
$\ket{\Uparrow_2}=\ot_{k=2}^N \ket{\uparrow}_k$. Obviously, we have
\begin{equation}\label{Eket}
\bra{\Downarrow_1}B(\lambda_n)\cdots B(\lambda_1)\ket{\Uparrow}=
E_{21}(\lambda_1,\dots,\lambda_n)\ket{\Uparrow_2}.
\end{equation}
Taking into account that
\begin{equation}
A_2(\lambda)\ket{\Uparrow_2}=\prod_{k=2}^{N} a(\lambda,\nu_k)\ket{\Uparrow_2}
\end{equation}
we can use \eqref{AB} to reduce RHS of \eqref{Eket} in terms of $B_2$'s only
(applied to vector $\ket{\Uparrow_2}$). The result reads
\begin{multline}\label{key}
E_{21}(\lambda_1,\dots,\lambda_n)\ket{\Uparrow_2}
=c
\sum_{\alpha=1}^{n}
\prod_{\substack{\beta=1 \\ \beta\ne\alpha}}^{n} b(\lambda_\beta,\nu_1)
\prod_{\substack{\beta=1\\ \beta\ne\alpha}}^{n}
f (\lambda_\alpha,\lambda_\beta)
\prod_{k=2}^{N}a(\lambda_\alpha,\nu_k)
\\ \times
B_2(\lambda_n)\cdots B_2(\lambda_{\alpha+1})
B_2(\lambda_{\alpha-1}) \cdots B_2(\lambda_1)
\ket{\Uparrow_2}.
\end{multline}
To get a hint about how this formula can be derived, it is sufficient to
look at the term $\alpha=n$ in \eqref{E21} which is proportional to
$A_2(\lambda_n)B_2(\lambda_{n-1})\cdots B(\lambda_1)$.
This is the only term which
contributes to the $\alpha=n$ term of \eqref{key}
(containing $B(\lambda_{n-1})\cdots B(\lambda_1)$) after applying
commutation relation \eqref{AB}; moreover, only the first term in RHS of
\eqref{AB} contributes to the  $\alpha=n$ term in \eqref{key}.
The remaining terms in \eqref{key} are just
due to the total symmetry with respect to permutations of $\lambda$'s.

Formulae \eqref{Eket} and \eqref{key}  express a vector
containing $B$'s in terms of
vectors containing $B_2$'s, so they  can be seen as a
recurrence relation with respect to $N$, the number of lattice sites.
Choosing a specific value of $n$ (the number of $B$'s)
and specifying also the component of this vector, one can
obtain recurrence relations for some important `scalar' quantities, such as
partition function and EFP of the domain-wall six-vertex model.
In turn, these recurrence relations can be explicitly solved.

\subsection{Recurrence relation for $Z_N$}

As an illustration of the approach
let us consider here how the recurrence relation for
$Z_N$ can be derived and solved. The recurrence relation emerges when
$n$ is specified to the value
$n=N$, and taking the scalar product of \eqref{Eket}
with the vector $\bra{\Downarrow_2}$. In this case relation \eqref{key} gives
\begin{equation}\label{recZN}
Z_N= c \sum_{\alpha=1}^{N}
\prod_{\substack{\beta=1 \\ \beta\ne\alpha}}^{N} b(\lambda_\beta,\nu_1)
\prod_{\substack{\beta=1\\ \beta\ne\alpha}}^{N}
f (\lambda_\alpha,\lambda_\beta)
\prod_{k=2}^{N}a(\lambda_\alpha,\nu_k)
Z_{N-1}[\lambda_\alpha;\nu_1].
\end{equation}
Here $Z_{N-1}[\lambda_\alpha;\nu_1]$ denotes the partition function of the
domain-wall six-vertex model on $(N-1)\times(N-1)$ lattice, with the sets
of $\lambda$'s and $\nu$'s such that they
do not contain $\lambda_\alpha$ and $\nu_1$,
namely, they are
$\lambda_1,\dots,\lambda_{\alpha-1},\lambda_{\alpha+1},\dots,\lambda_N$, and
$\nu_2,\dots,\nu_N$, respectively (in other words,
the square brackets indicate independence of
these variables, in comparison with the `original' sets
$\lambda_1,\dots,\lambda_N$ and $\nu_1,\dots,\nu_N$).

Relation \eqref{recZN} obviously represents a recurrence
relation for the partition function with respect to the size of the lattice.
The initial condition to the recurrence is $Z_1=c$.
It is to be emphasized that in \eqref{recZN} values of
$\lambda$'s and $\nu$'s are completely arbitrary.

In order to prove  that Izergin-Korepin formula indeed
solves \eqref{recZN}, one can just
substitute \eqref{ZN} into  both sides of this relation, and verify
whether this is an identity or not.
After substituting  \eqref{ZN} into \eqref{recZN}, and cancelling
many factors, one is left with
\begin{equation}\label{simple}
\det\caM =
\frac{\prod_{k=2}^{N}d(\nu_1,\nu_k)}{\prod_{\alpha=1}^{N}a(\lambda_\alpha,\nu_1)}
\sum_{\alpha=1}^{N} (-1)^{\alpha-1} g(\lambda_\alpha) \det\caM_{[\alpha;1]}.
\end{equation}
Here $\caM_{[\alpha;1]}$ denotes the $(N-1)\times(N-1)$ matrix obtained from
matrix $\caM$ \eqref{matT} by removing $\alpha$-th row and the first column.
The function $g(\lambda)$ is defined by
\begin{equation}\label{gNdef}
g(\lambda):=
\frac{\prod_{\alpha=1}^{N}
e(\lambda_\alpha,\lambda)}{\prod_{k=1}^{N}b(\lambda,\nu_k)},
\end{equation}
where
\begin{equation}\label{efunc}
e(\lambda,\lambda'):=\sin(\lambda-\lambda'+2\eta).
\end{equation}
It is useful to note that the sum in \eqref{simple} can be written as
the determinant
\begin{equation}
\begin{vmatrix}
g(\lambda_1) & \varphi(\lambda_1,\nu_2) & \dots & \varphi(\lambda_1,\nu_N) \\
\hdotsfor{4} \\
g(\lambda_N) & \varphi(\lambda_N,\nu_2) & \dots & \varphi(\lambda_N,\nu_N)
\end{vmatrix}.
\end{equation}
The main point is that for each value $\lambda=\lambda_\alpha$
($\alpha=1,\dots,N$) the function $g(\lambda)=
g(\lambda;\lambda_1,\dots,\lambda_N;\nu_1,\dots,\nu_N)$, where
$\lambda_1,\dots,\lambda_N$ and $\nu$'s are to be regarded as parameters,
can be represented as follows
\begin{equation}\label{gNsum}
g(\lambda_\alpha)=\sum_{k=1}^{N} \varphi(\lambda_\alpha,\nu_k)
\frac{\prod_{\alpha=1}^{N}a(\lambda_\alpha,\nu_k)}
{\prod_{\substack{j=1\\ j\ne k}}^{N}d(\nu_k,\nu_j)}.
\end{equation}
Noting that the ratio of the two products here does not
depend on $\alpha=1,\dots,N$,
it is easy to see that relation \eqref{simple} is indeed fulfilled.

The only nontrivial point in this derivation is identity \eqref{gNsum} which
can be proven by various methods (e.g., by induction). Let us
 mention here a hint (even if not completely rigourous)
on how such kind of relations can be deduced: at values $\lambda=\nu_k+\eta$
($k=1,\dots,N$) function $g(\lambda)$ has simple poles; the sum in
RHS of
\eqref{gNsum} is nothing but the sum over these poles, in analogy with
meromorphic function expansion.

\section{Calculation of EFP}

\subsection{Recurrence relation}

Let us denote
\begin{equation}
\wt F_N^{(r,s)}:=Z_N^{}  F_N^{(r,s)},
\end{equation}
that is, we will consider temporarily the sole matrix element of
\eqref{EFP}, or the `numerator' of the correlation function.
Choosing in \eqref{Eket} $n=r$ and taking the scalar product
of this vector with
$\bra{\Downarrow_2}B_2(\lambda_N)\cdots B_2(\lambda_{r+1})$, one
can easily see that \eqref{key} implies the following recurrence relation
\begin{multline}\label{recEFP}
\wt F_N^{(r,s)}= \prod_{\beta=r+1}^{N} a(\lambda_\beta,\nu_1)\,
c \sum_{\alpha=1}^{r}
\prod_{\substack{\beta=1 \\ \beta\ne\alpha}}^{r} b(\lambda_\beta,\nu_1)
\prod_{\substack{\beta=1\\ \beta\ne\alpha}}^{r}
f (\lambda_\alpha,\lambda_\beta)
\prod_{k=2}^{N}a(\lambda_\alpha,\nu_k)
\\ \times
\wt F_{N-1}^{(r-1,s-1)}[\lambda_\alpha;\nu_1].
\end{multline}
As above, the `dependence' on some $\lambda_\alpha$ and $\nu_k$ enclosed
in brackets indicates that the sets of $\lambda$'s and $\nu$'s on which
the quantity depends, has $N-1$ elements each, with $\lambda_\alpha$
and $\nu_k$ missing in the corresponding set (i.e., brackets indicate
no dependence on these variables).

In the case $s=1$ EFP
describes the `boundary' polarization, which
was considered in \cite{BPZ-02}. For all $r=1,\dots,N$, we have
\begin{equation}\label{FeqZ}
\wt
F_{N-1}^{(r-1,0)}[\lambda_\alpha;\nu_1]=Z_{N-1}^{}[\lambda_\alpha;\nu_1]
\end{equation}
and therefore we can just plug Izergin-Korepin formula into RHS of
\eqref{recEFP} in order to obtain EFP at $s=1$. Indeed, taking into account that
\begin{multline}\label{ratio}
\frac{Z_{N-1}[\lambda_\alpha,\nu_1]}{Z_N}=
\frac{(-1)^{\alpha-1}}{a(\lambda_\alpha,\nu_1)b(\lambda_\alpha,\nu_1)}
\prod_{\substack{\beta=1\\ \beta\ne\alpha}}^{N}
\frac{d(\lambda_\beta,\lambda_\alpha)}{a(\lambda_\beta,\nu_1)b(\lambda_\beta,\nu_1)}
\prod_{k=2}^{N}\frac{d(\nu_1,\nu_k)}{a(\lambda_\alpha,\nu_k)b(\lambda_\alpha,\nu_k)}
\\ \times
\frac{\det\caM_{[\alpha;1]}}{\det\caM}
\end{multline}
we arrive at the expression
\begin{equation}\label{EFP1}
F_{N}^{(r,1)}=\frac{1}{\det\caM}\cdot\frac{\prod_{k=2}^{N}d(\nu_1,\nu_k)}
{\prod_{\alpha=1}^{r}a(\lambda_\alpha,\nu_1)
\prod_{\alpha=r+1}^{N}b(\lambda_\alpha,\nu_1)}
\sum_{\alpha=1}^{r} (-1)^{\alpha-1} g_r(\lambda_\alpha)\,
\det\caM_{[\alpha;1]}.
\end{equation}
Here the function
$g_r(\lambda):=g_r(\lambda;\lambda_1,\dots,\lambda_N;\nu_1,\dots,\nu_N)$,
is given by
\begin{equation}\label{gr}
g_r(\lambda):= \frac{\prod_{\alpha=r+1}^{N} d(\lambda_\alpha,\lambda)
\prod_{\alpha=1}^{r} e(\lambda_\alpha,\lambda)
}{\prod_{k=1}^{N}b(\lambda,\nu_k)},
\end{equation}
where function $e(\lambda,\lambda')$ is defined in \eqref{efunc}. For
$r=N$ function $g_r(\lambda)$ is just function $g(\lambda)$ defined in
\eqref{gNdef}.

\subsection{The cases $s=2$ and $s=3$} Using \eqref{recEFP} and \eqref{EFP1} we can
derive EFP in the case of $s=2$. Indeed, for $\alpha=1,\dots,r$, using \eqref{EFP1},
we can write
\begin{multline}\label{smaller}
F_{N-1}^{(r-1,1)}[\lambda_\alpha;\nu_1]=
\frac{1}{\det\caM_{[\alpha;1]}}
\prod_{k=3}^{N}d(\nu_2,\nu_k)
\prod_{\substack{\beta=1\\ \beta\ne\alpha}}^{r}\frac{1}{a(\lambda_\beta,\nu_2)}
\prod_{\beta=r+1}^{N}\frac{1}{b(\lambda_\beta,\nu_2)}
\\ \times
\sum_{\substack{\beta=1\\ \beta\ne\alpha}}^{r} (-1)^{\beta-1+\chi(\beta,\alpha)}
\frac{b(\lambda_\beta,\nu_1)}{e(\lambda_\alpha,\lambda_\beta)}\,g_r(\lambda_\beta)\,
\det\caM_{[\alpha,\beta;1,2]}.
\end{multline}
Here $\chi(\beta,\alpha)=1$ if $\beta>\alpha$, and
$\chi(\beta,\alpha)=0$ otherwise. Substituting in \eqref{smaller} the
expression for $\det\caM_{[\alpha;1]}$ which follows from
relation \eqref{ratio}, and switching to the non-normalized quantity, we get
\begin{multline}
\wt F_{N-1}^{(r-1,1)}[\lambda_\alpha;\nu_1]=
\frac{Z_N}{\det\caM}
\prod_{\substack{\beta=1\\ \beta\ne\alpha}}^{N}
\frac{d(\lambda_\beta,\lambda_\alpha)}{a(\lambda_\beta,\nu_1)b(\lambda_\beta,\nu_1)}
\prod_{k=2}^{N}\frac{d(\nu_1,\nu_k)}{a(\lambda_\alpha,\nu_k)b(\lambda_\alpha,\nu_k)}
\\ \times
\prod_{k=3}^{N}d(\nu_2,\nu_k)
\prod_{\substack{\beta=1\\ \beta\ne\alpha}}^{r}\frac{1}{a(\lambda_\beta,\nu_2)}
\prod_{\beta=r+1}^{N}\frac{1}{b(\lambda_\beta,\nu_2)}
\\ \times
\sum_{\substack{\beta=1\\ \beta\ne\alpha}}^{r} (-1)^{\beta-1+\chi(\beta,\alpha)}
\frac{b(\lambda_\beta,\nu_1)}{e(\lambda_\alpha,\lambda_\beta)}g_r(\lambda_\beta)
\det\caM_{[\alpha,\beta;1,2]}.
\end{multline}
Plugging this expression into the recurrence relation, after all
cancellations, we finally find
\begin{multline}\label{EFP2}
F_N^{(r,2)}=\frac{1}{\det\caM}\cdot
\frac{\prod_{k=2}^{N}d(\nu_1,\nu_k)\prod_{k=3}^{N}d(\nu_2,\nu_k)}
{\prod_{\alpha=1}^{r}a(\lambda_\alpha,\nu_1) a(\lambda_\alpha,\nu_2)
\prod_{\alpha=r+1}^{N}b(\lambda_\alpha,\nu_1)b(\lambda_\alpha,\nu_2)
}
\\ \times
\sum_{\alpha=1}^{r}
\sum_{\substack{\beta=1\\ \beta\ne\alpha}}^{r}
(-1)^{\alpha+\beta+\chi(\beta,\alpha)}
\frac{a(\lambda_\alpha,\nu_2)b(\lambda_\beta,\nu_1)}
{e(\lambda_\alpha,\lambda_\beta)}g_r(\lambda_\alpha)g_r(\lambda_\beta)
\det\caM_{[\alpha,\beta;1,2]}.
\end{multline}

Expression \eqref{EFP2} can be further used to find EFP
for $s=3$, by repeating the procedure just explained.
We quote here only the result,
\begin{multline}\label{EFP3}
F_N^{(r,3)}=\frac{1}{\det\caM}
\prod_{j=1}^{3}
\frac{\prod_{k=j+1}^{N}d(\nu_j,\nu_k)}
{\prod_{\alpha=1}^{r}a(\lambda_\alpha,\nu_j)
\prod_{\alpha=r+1}^{N}b(\lambda_\alpha,\nu_j)}
\\ \times
\sum_{\alpha=1}^{r} \sum_{\substack{\beta=1\\ \beta\ne\alpha}}^{r}
\sum_{\substack{\gamma=1\\ \gamma\ne\beta,\alpha}}^{r}
(-1)^{\alpha+\beta+\gamma+1+\chi(\gamma,\alpha)+\chi(\gamma,\beta)
+\chi(\beta,\alpha)}
g_r(\lambda_\alpha)g_r(\lambda_\beta)g_r(\lambda_\gamma)
\\ \times
\frac{a(\lambda_\alpha,\nu_2) a(\lambda_\alpha,\nu_3) a(\lambda_\beta,\nu_3)
b(\lambda_\beta,\nu_1) b(\lambda_\gamma,\nu_1) b(\lambda_\gamma,\nu_2)}
{e(\lambda_\alpha,\lambda_\beta)
e(\lambda_\alpha,\lambda_\gamma)
e(\lambda_\beta,\lambda_\gamma)}
\\  \times
\det\caM_{[\alpha,\beta,\gamma;1,2,3]}\,.
\end{multline}

\subsection{Result for generic $s$}

Inspecting formulae \eqref{EFP1}, \eqref{EFP2} and \eqref{EFP3} it is rather
straightforward to guess the result for generic values of $s$.
The following expression for EFP is valid:
\begin{multline}\label{EFPs}
F_N^{(r,s)}=
\frac{1}{\det\caM}
\prod_{j=1}^{s}\frac{\prod_{k=j+1}^{N}d(\nu_j,\nu_k)}
{\prod_{\beta=1}^{r}a(\lambda_\beta,\nu_j)\prod_{\beta=r+1}^{N}
b(\lambda_\beta,\nu_j)}
\\ \times
\sum_{\alpha_1=1}^{r} \sum_{\substack{\alpha_2=1\\ \alpha_2\ne\alpha_1}}^{r}
\cdots
\sum_{\substack{\alpha_s=1\\ \alpha_s\ne\alpha_1,\dots,\alpha_{s-1}}}^{r}
(-1)^{s+\sum_{k=1}^{s}\alpha_k+\sum_{1\leq j<k\leq s}\chi(\alpha_k,\alpha_j)}
\prod_{j=1}^{s} g_r(\lambda_{\alpha_j})
\\ \times
\prod_{1\leq j<k\leq s}^{}
\frac{a(\lambda_{\alpha_j},\nu_k)b(\lambda_{\alpha_k},\nu_j)}
{e(\lambda_{\alpha_j},\lambda_{\alpha_k})}
\det\caM_{[\alpha_1,\dots,\alpha_s;1,\dots,s]}\,.
\end{multline}
It can be shown directly that formula \eqref{EFPs} solves recurrence
relation \eqref{recEFP}. The calculation in fact repeats the one above
for the case of $s=2$.

Let us focus on
writing EFP for the case of $(N-1)\times (N-1)$ lattice, namely,
we are interested in $F_{N-1}^{(r-1,s-1)}[\lambda_{\alpha_1},\nu_1]$.
To be more precise, in adapting expression
\eqref{EFPs} to this case, apart from using the fact
that the sum is now $(s-1)$-fold, it is important also to take
into account that the function
$g_r(\lambda)$ implicitly depends on $N$, so that switching from
$N$ to $(N-1)$ implies some extra factor.
In all, after all these preparations, which are mostly devoted to
fit the notations, the expression which has to be
substituted into RHS of \eqref{recEFP} reads:
\begin{multline}\label{EFPsm1}
F_{N-1}^{(r-1,s-1)}[\lambda_{\alpha_1},\nu_1]
=\frac{1}{\det\caM_{[\alpha_1;1]}}
\prod_{j=2}^{s}\frac{\prod_{k=j+1}^{N}d(\nu_j,\nu_k)}
{\prod_{\substack{\beta=1 \\ \beta\ne \alpha_1}}^{r}a(\lambda_{\beta},\nu_j)
\prod_{\beta=r+1}^{N}b(\lambda_{\beta},\nu_j)}
\\ \times
\sum_{\substack{\alpha_2=1 \\ \alpha_2\ne\alpha_1}}^{r}\cdots
\sum_{\substack{\alpha_s=1 \\ \alpha_s\ne\alpha_1,\dots,\alpha_{s-1}}}^{r}
(-1)^{s-1+\sum_{k=2}^s \alpha_k+\sum_{1\leq j<k\leq s}\chi(\alpha_k,\alpha_j)}
\\ \times
\prod_{j=1}^s\frac{g_r(\lambda_{\alpha_j}) b(\lambda_{\alpha_j},\nu_1)}
{e(\lambda_{\alpha_1},\lambda_{\alpha_j})}
\prod_{2\leq j<k\leq s}\frac{a(\lambda_{\alpha_j},\nu_k)b(\lambda_{\alpha_k},\nu_j)}
{e(\lambda_{\alpha_j},\lambda_{\alpha_k})}
\\ \times
\det\caM_{[\alpha_1,\dots,\alpha_s;1,\dots,s]}\,.
\end{multline}
Again, as in the case of $s=2$ considered above, one can use
relation \eqref{ratio} to eliminate $\det\caM_{[\alpha_1;1]}$ in
favour of $\det\caM$ in \eqref{EFPsm1}. With this point taken into account,
cancellation of various factors shows that
recurrence relation \eqref{recEFP} is indeed fulfilled.

Apart from showing that \eqref{EFPs} is a solution to \eqref{recEFP},
it is also useful to mention that the corresponding `initial
conditions' are also satisfied. Indeed, for
$s\leq r$, such a condition is just the generalization of condition \eqref{FeqZ},
namely
\begin{equation}
\tilde F_{N-s}^{(r-s,0)}
[\lambda_{\alpha_1},\dots,\lambda_{\alpha_s};\nu_1,\dots,\nu_s]=
Z_{N-s}[\lambda_{\alpha_1},\dots,\lambda_{\alpha_s};\nu_1,\dots,\nu_s].
\end{equation}
For $s>r$, the corresponding condition is
\begin{equation}
F_{N-r}^{(0,s-r)}[\lambda_{\alpha_1},\dots,\lambda_{\alpha_r};\nu_1,\dots,\nu_r]=0.
\end{equation}
Note, that this relation implies the relation $F_N^{(r,s)}=0$
which must hold whenever $s>r$, and which
follows directly from the definition of EFP \eqref{EFP}.

\section{EFP in the homogeneous limit}

\subsection{The procedure}

The homogeneous limit can be performed along the lines of papers
\cites{ICK-92,CP-05c}. We start with
writing
\begin{equation}
\lambda_\alpha=\lambda+\xi_\alpha,
\end{equation}
where $\xi$'s will be set
equal to zero in the limit (as well as $\nu$'s). Keeping
$\xi$'s nonzero (and different from each other), and
using the fact that for a function $f(x)$, regular
near $x=\lambda$, one has $\exp(\xi\partial_\eps) f(\lambda+\eps)|_{\eps=0}
= f(\lambda+\xi)$, we can bring \eqref{EFPs} into a form
which involves some determinant:
\begin{multline}\label{EFPdet}
F_N^{(r,s)}=
\frac{1}{\det\caM}
\prod_{j=1}^{s}\frac{\prod_{k=j+1}^{N}d(\nu_j,\nu_k)}
{\prod_{\beta=1}^{r}a(\lambda_\beta,\nu_j)\prod_{\beta=r+1}^{N}
b(\lambda_\beta,\nu_j)}
\\ \times
\begin{vmatrix}
\exp(\xi_1\partial_{\eps_1}) & \dots & \exp(\xi_1\partial_{\eps_s})
& \varphi(\lambda_1,\nu_{s+1}) & \dots & \varphi(\lambda_1,\nu_{N})
\\
\hdotsfor{6}
\\
\exp(\xi_N\partial_{\eps_1}) & \dots & \exp(\xi_N\partial_{\eps_s})
& \varphi(\lambda_N,\nu_{s+1}) & \dots & \varphi(\lambda_N,\nu_{N})
\end{vmatrix}
\\ \times
\prod_{j=1}^{s} g_r(\lambda+\eps_j)
\prod_{1\leq j<k\leq s}^{}
\frac{a(\lambda+\eps_j,\nu_k)b(\lambda+\eps_k,\nu_j)}
{e(\lambda+\eps_j,\lambda+\eps_k)}\Bigg|_{\eps_1=\ldots=\eps_s=0}.
\end{multline}
It is to be emphasised that this expression  is valid for
the inhomogeneous model (no homogeneous limit yet); it represents
an equivalent way of writing the multiple sum in \eqref{EFPs}.

Let us now perform the limit $\nu_1\to 0$, $\nu_2\to 0$, $\ldots$,
$\nu_N\to 0$, in this order, at each stage keeping the contribution of
leading order in the corresponding variable;
next we do the same with $\xi$'s, in the order $\xi_1\to 0$,
$\xi_2\to 0$, $\ldots$, $\xi_N\to 0$, again keeping only the contributions
of leading order. The first line of \eqref{EFPdet} gives
\begin{equation}
\frac{(-1)^{\frac{(N-s-1)(N-s)}{2}}\prod_{j=1}^{s} (N-j)!}
{a^{rs} b^{(N-r)s}\det\caN}
\cdot
\frac{1}{\frac{\nu_{s+2}}{1!}\cdot
\frac{\nu_{s+3}^2}{2!}\cdots \frac{\nu_{N}^{N-s-1}}{(N-s-1)!}}
\cdot
\frac{1}{\frac{\xi_{2}}{1!}\cdot\frac{\xi_{3}^2}{2!}\cdots
\frac{\xi_{N}^{N-1}}{(N-1)!}},
\end{equation}
where $a:=a(\lambda,0)$, $b:=b(\lambda,0)$. The
second line of \eqref{EFPdet} gives
\begin{multline}\label{secondline}
(-1)^{\frac{(N-s-1)(N-s)}{2}}
\begin{vmatrix}
1& \dots &  1& \varphi(\lambda)& \dots & \partial_\lambda^{N-s-1}\varphi(\lambda)
\\
\partial_{\eps_1} & \dots & \partial_{\eps_s} &
\partial_\lambda\varphi(\lambda) & \dots &
\partial_\lambda^{N-s}\varphi(\lambda)
\\ \hdotsfor{6} \\
\partial_{\eps_1}^{N-1} & \dots & \partial_{\eps_s}^{N-1} &
\partial_\lambda^{N-1}\varphi(\lambda) & \dots &
\partial_\lambda^{2N-s-2}\varphi(\lambda)
\end{vmatrix}
\\ \times
\left(\frac{\nu_{s+2}}{1!}\cdot
\frac{\nu_{s+3}^2}{2!}\cdots \frac{\nu_{N}^{N-s-1}}{(N-s-1)!}
\right)
\cdot\left(
\frac{\xi_{2}}{1!}\cdot\frac{\xi_{3}^2}{2!}\cdots
\frac{\xi_{N}^{N-1}}{(N-1)!}\right).
\end{multline}
The terms in the third line of \eqref{EFPdet} give
\begin{equation}
\prod_{j=1}^{s}
\frac{[-\sin(\eps_j-2\eta)]^r(-\sin\eps_j)^{N-r}}{[\sin(\eps_j+\lambda-\eta)]^N}
\prod_{1\leq j<k \leq s}^{} \frac{\sin(\eps_j+\lambda+\eta)
\sin(\eps_k+\lambda-\eta)}{\sin(\eps_j-\eps_k+2\eta)}.
\end{equation}
In all, after all cancellations, EFP in the homogeneous model reads
\begin{multline}\label{homEFP}
F_N^{(r,s)}=\frac{(-1)^s\prod_{j=1}^{s}(N-j)!}{a^{rs}
b^{(N-r)s}\det\caN}
\begin{vmatrix}
\varphi(\lambda)& \dots & \partial_\lambda^{N-s-1}\varphi(\lambda)& 1& \dots &  1
\\
\partial_\lambda\varphi(\lambda) & \dots & \partial_\lambda^{N-s}\varphi(\lambda)
& \partial_{\eps_1} & \dots & \partial_{\eps_s}
\\ \hdotsfor{6} \\
\partial_\lambda^{N-1}\varphi(\lambda) & \dots &
\partial_\lambda^{2N-s-2}\varphi(\lambda)
& \partial_{\eps_1}^{N-1} & \dots & \partial_{\eps_s}^{N-1}
\end{vmatrix}
\\ \times
\prod_{j=1}^{s}
\frac{(\sin\eps_j)^{N-r}[\sin(\eps_j-2\eta)]^r}{[\sin(\eps_j+\lambda-\eta)]^N}
\\ \times
\prod_{1\leq j<k \leq s}^{} \frac{\sin(\eps_j+\lambda+\eta)
\sin(\eps_k+\lambda-\eta)}{\sin(\eps_j-\eps_k+2\eta)}
\Bigg|_{\eps_1=\ldots=\eps_s=0}.
\end{multline}
Note, that in the determinant here  we have changed the order of columns,
in comparison with formulae \eqref{EFPdet} or
\eqref{secondline}.

\subsection{Orthogonal polynomials representation}

Formula \eqref{homEFP} for EFP of the homogenous model, involving
$N\times N$ determinant, can be also represented
in terms of some $s\times s$ determinant.
Such an equivalent representation can be further used
to obtain a multiple integral representation
(having $s$ integrations), similar to those
for correlation functions of quantum spin chains
\cites{JM-95,KMT-00,BKS-03,BJMST-06}.

The derivation of the $s\times s$ determinant representation from
\eqref{homEFP} is based on the following facts. Let $\{P_n(x)\}_{n=0}^\infty$ be a
set of orthogonal polynomials,
\begin{equation}\label{ortho}
\int   P_{n}(x) P_{m}(x) \mu(x)\,\rmd x
=h_{n} \delta_{nm} ,
\end{equation}
where the integration domain is assumed over the real axis, the weight
$\mu(x)$ is real nonnegative, and we choose $h_n$'s such that
$P_n(x)$'s are monic (i.e., the leading coefficient is equal
to one, $P_n(x)=x^n + \dots$). Let $c_n$ denote $n$-th
moment of the weight $\mu(x)$,
\begin{equation}
c_n =\int x^n \mu(x)\, \rmd x
\qquad (n=0,1,\ldots).
\end{equation}
The orthogonality condition \eqref{ortho} and
standard properties of determinants allow one to prove that
\begin{equation}
\begin{vmatrix}
c_0&c_1& \dots &c_{n-1} \\
c_1& c_2&  \dots & c_{n}\\
\hdotsfor{4}  \\
c_{n-1} & c_{n} & \dots & c_{2n-2}
\end{vmatrix}
= h_0 h_1\cdots h_{n-1}.
\end{equation}
More generally, the following formula is valid
\begin{multline}\label{detdet}
\begin{vmatrix}
c_0&c_1& \dots &c_{n-k-1}  & 1 & 1 & \dots & 1 \\
c_1&c_2& \dots &c_{n-k} & x_1 & x_2 &\dots &x_k \\
\hdotsfor{8}  \\
c_{n-1}& c_{n} &\dots & c_{2n-k-1} & x_1^{n-1} & x_2^{n-1} &\dots &x_k^{n-1}
\end{vmatrix}
\\
=h_0 \cdots h_{n-k-1}
\begin{vmatrix}
P_{n-k}(x_1) & \dots & P_{n-k}(x_k)\\
\hdotsfor{3} \\
P_{n-1}(x_1) & \dots & P_{n-1}(x_k)
\end{vmatrix}.
\end{multline}
The proof of this formula can be given by the methods explained in \cite{S-75}.
We shall use \eqref{detdet} with $n=N$ and $k=s$.

To apply \eqref{detdet} to expression \eqref{homEFP} we need first to
identify the
orthogonality weight. Since in our case, by definition,
$c_n=\partial_\lambda^n\varphi(\lambda)$, the weight
can be obtained by representing the function $\varphi(\lambda)$ via
Laplace transform:
\begin{equation}
\varphi(\lambda)=\int   \rme^{\lambda x}\Phi(x) \,\rmd x.
\end{equation}
Here function
$\Phi(x)=\Phi(x;\eta)$ is independent of $\lambda$, but depends on $\eta$
as a parameter (indeed, in our case $\mu(x)=\mu(x;\lambda,\eta)$ where
$\lambda$ and $\eta$ are to be considered as parameters,
and $\mu(x)=\rme^{\lambda x} \Phi(x)$).

The explicit form of function $\Phi(x)$ depends on the particular choice
of the physical regime in the six-vertex model. For example, let us
consider the so-called disordered regime ($|\Delta|<1$). In this case
$\lambda$ and $\eta$ in our standard parametrization of weights
$a=\sin(\lambda+\eta)$,
$b=\sin(\lambda-\eta)$, and $c=\sin(2\eta)$ are both real and
satisfy $0<\eta<\pi/2$ and $\eta<\lambda<\pi-\eta$.
Function $\Phi(x)$ can be found from the formula:
\begin{equation}\label{int}
\int_{-\infty}^{\infty} \rme^{x(\lambda-\pi/2)}
\frac{\sinh(\eta x)}{\sinh(\pi x/2)}\; \rmd x
=\frac{\sin(2\eta)}{\sin(\lambda-\eta)\sin(\lambda+\eta)}.
\end{equation}
The other two regimes, the ferroelectric ($\Delta>1$) and anti-ferroelectric
($\Delta<-1$) can be approached when $\lambda$ and $\eta$ are complex, satisfying
$\Re \lambda =\Re\eta =0$ or $\Re \lambda =\Re\eta =\pi/2$, respectively.
In fact, for these regimes the weight $\mu(x)$ can be found from \eqref{int}
using the proper analytical
continuation in the parameters $\lambda$ and $\eta$.
The integral in RHS of \eqref{int} in these cases is replaced  by a
sum coming from the simple poles of the integrand, or, in other words,
the corresponding measure $\mu(x)\rmd x$ turns out to be
discrete (see \cite{Zj-00} for explicit formulae). For our construction below
the actual choice of the regime, and hence the explicit expression
for the orthogonality weight $\mu(x)$, is irrelevant since only
the corresponding orthogonal polynomials will enter the formulae explicitly
(for instance,
one can think of the case of the disordered regime; the resulting
expressions in terms of $P_n(x)$'s for other regimes
turn out to be essentially the same).

To make the use of \eqref{detdet} convenient in the framework of
representation \eqref{homEFP}, we define, following \cite{CP-05c},
the functions
\begin{equation}\label{omegas}
\omega(\epsilon)=\frac{\sin(\lambda+\eta)}{\sin(\lambda-\eta)}\,
\frac{\sin\eps}{\sin(\eps-2\eta)},\qquad
\tilde\omega(\epsilon)=\frac{\sin(\lambda-\eta)}{\sin(\lambda+\eta)}\,
\frac{\sin\eps}{\sin(\eps+2\eta)}.
\end{equation}
They are functions of $\eps$, with $\lambda$ and $\eta$ regarded as parameters,
as indicated in the notations (below we omit the dependence
on $\lambda$ and $\eta$ whenever possible). We also define
\begin{equation}
\varrho(\epsilon)=\frac{\sin(\lambda-\eta)}{\sin(2\eta)}\,
\frac{\sin(\eps-2\eta)}{\sin(\eps+\lambda-\eta)},
\qquad
\tilde\varrho(\epsilon)=\frac{\sin(\lambda+\eta)}{\sin(2\eta)}\,
\frac{\sin(\eps+2\eta)}{\sin(\eps+\lambda+\eta)}.
\end{equation}
There are useful relations
\begin{equation}\label{go}
\varrho(\eps)=\frac{1}{\omega(\eps)-1},\qquad
\tilde\varrho(\eps)=\frac{1}{1-\tilde\omega(\eps)}.
\end{equation}
Noting that
\begin{equation}
\frac{\sin(\eps_1+\lambda+\eta)\sin(\eps_2+\lambda-\eta)}
{\sin(\eps_1-\eps_2+2\eta)}
=\frac{1}{\varphi \tilde\varrho(\eps_1) \varrho(\eps_2)}\;
\frac{1}{\tilde\omega(\eps_1)\omega(\eps_2)-1}.
\end{equation}
and defining
\begin{equation}\label{Knx}
K_n(x)= \frac{n!\,\varphi^{n+1}}{h_n}\; P_n(x)
\end{equation}
we have, in virtue \eqref{detdet}, the following orthogonal polynomials
representation:
\begin{multline}\label{orthEFP}
F_N^{(r,s)}=(-1)^s
\begin{vmatrix}
K_{N-s}(\partial_{\eps_1}) &\hdots & K_{N-s}(\partial_{\eps_s})\\
\hdotsfor{3} \\
K_{N-1}(\partial_{\eps_1}) &\hdots & K_{N-1}(\partial_{\eps_s})
\end{vmatrix}
\prod_{j=1}^{s}\left\{ [\omega(\eps_j)]^{N-r}[\varrho(\eps_j)]^N \right\}
\\ \times
\prod_{1\leq j<k \leq s}^{} \frac{1}
{\tilde\varrho(\eps_j)\varrho(\eps_k)[\tilde\omega(\eps_j)\omega(\eps_k)-1]}
\Bigg|_{\eps_1=\ldots=\eps_s=0}.
\end{multline}

\subsection{Multiple integral representation}

Let $H_N^{(r)}$ denotes the probability
of having the  $c$-weight vertex of the first row (from the top of the lattice)
at $r$-th position (from the right). This quantity can be related to
EFP at $s=1$, which describes in this case essentially the
one-point correlation function (polarization) at the boundary, via
$H_N^{(r)}=F_N^{(r,1)}-F_N^{(r-1,1)}$
(we refer for details to \cite{BPZ-02} where
$F_N^{(r,1)}$ was denoted as $G_N^{(r)}$). Using \eqref{orthEFP} and taking into
account the first relation in \eqref{go}, we have (see also \cite{CP-05c})
\begin{equation}\label{H_N}
H_N^{(r)}=
K_{N-1}(\partial_\eps)\,\frac{[\omega(\eps)]^{N-r}}{[\omega(\eps)-1]^{N-1}}
\bigg|_{\epsilon=0}.
\end{equation}
Let us define the generating function:
\begin{equation}\label{hNz}
h_N(z)=\sum_{r=1}^{N} H_N^{(r)}z^{r-1}.
\end{equation}
The key identity, for the derivation of a multiple integral representation
for EFP, is
\begin{equation}\label{claim}
K_{N-1}(\partial_\eps)\, f(\omega(\eps))\Big|_{\eps=0}=
\frac{1}{2\pi i}\oint_{C_0}^{} \frac{(z-1)^{N-1}}{z^N} h_N(z) f(z)\, \rmd z.
\end{equation}
Here $f(z)$ is some function, regular near $z=0$, and
$C_0$ is a simple closed counterclockwise countour around the origin.

To prove this identity, we note first that since $f(z)$ is assumed regular at
$z=0$, one can further think of it as a polynomial
of degree $(N-1)$, since higher powers in $z$ do not contribute to either
side of
\eqref{claim} (note that $\omega(\eps)\to 0$ as $\eps\to 0$).
It therefore suffices to prove \eqref{claim} for $f(z)$ being a monomial.
Next, let us define auxiliary quantities
\begin{equation}\label{T_N}
V_N^{(p)}=K_{N-1}(\partial_\eps)\, [\omega(\eps)]^{N-p}\big|_{\epsilon=0},
\qquad (p=1,\dots,N).
\end{equation}
Evaluating the integral in RHS of \eqref{claim}, one finds that
\eqref{claim} is nothing but the relation
\begin{equation}\label{tAh}
\vec v =(-1)^{N-1}\, \caA \vec h,
\end{equation}
where the vectors $\vec v$ and $\vec h$ have components
$v_p=V_N^{(p)}$ and $h_r=H_N^{(r)}$, and matrix $\caA$ has entries
$\caA_{pr} = (-1)^{p-r}  \binom{N-1}{p-r}$.
Obviously,  matrix $\caA$ is lower-triangular, and moreover it can be
represented as
\begin{equation}\label{power}
\caA=(I-\caE)^{N-1},
\end{equation}
where $\caE$ denotes the lower triangular matrix, with entries
standing under the main diagonal equal to one and all other entries being zeroes,
i.e., $\caE_{pr}=\delta_{p,r+1}$. Inverting matrix $\caA$ in \eqref{tAh}
with the help of \eqref{power}, one arrives immediately to the expression for
$H_N^{(r)}$ in terms of $V_N^{(p)}$'s, which follows from
formula \eqref{H_N} and definition \eqref{T_N}. This proves identity \eqref{claim}.

As a result, applying identity \eqref{claim} to \eqref{orthEFP} and
taking into account relations \eqref{go} and definitions \eqref{omegas},
we obtain the following multiple integral representation for EFP:
\begin{multline}\label{MIR1}
F_N^{(r,s)} = \left(-\frac{1}{2\pi \rmi}\right)^s
\oint_{C_0}^{} \cdots \oint_{C_0}^{}
\begin{vmatrix}
\dfrac{h_{N}(z_1)}{z_1^{r}(z_1-1)}&  \hdots & \dfrac{h_{N}(z_s)}{z_s^{r}(z_s-1)} \\
\dfrac{h_{N-1}(z_1)}{z_1^{r-1}(z_1-1)^{2}}& \hdots
& \dfrac{h_{N-1}(z_s)}{z_s^{r-1}(z_s-1)^{2}}  \\
\hdotsfor{3} \\
\dfrac{h_{N-s+1}(z_1)}{z_1^{r-s+1}(z_1-1)^{s}}& \hdots
& \dfrac{h_{N-s+1}(z_s)}{z_s^{r-s+1}(z_s-1)^{s}}
\end{vmatrix}
\\ \times
\prod_{1\leq j < k \leq s}^{}
\frac{(\tilde z_j-1)(z_k-1)}{\tilde z_j z_k -1}\ \rmd z_1\cdots \rmd z_s.
\end{multline}
Here $\tilde z_j$'s are functions of the corresponding $z_j$'s,
\begin{equation}\label{tildez}
\tilde z_j = \frac{b^2z_j}{(a^2+b^2-c^2) z_j -a^2},
\end{equation}
where $a$, $b$, and $c$ are the homogeneous six-vertex model weights
(formula \eqref{tildez} follows from \eqref{omegas}
by expressing function
$\tilde\omega(\eps)$ in terms of $\omega(\eps)$ and
using $a=\sin(\lambda+\eta)$,
$b=\sin(\lambda-\eta)$, and $c=\sin(2\eta)$).

\section{More on the multiple integral representation}

\subsection{Preliminaries}

Formula \eqref{MIR1} gives EFP in terms of a multiple integral. This
integral representation can be transformed into some other forms which can be
useful for further study of EFP (e.g., for finding the limit shape).
Our aim in this section is to derive
some of these representations.

The first point to be mentioned in this respect is that
the integrand of \eqref{MIR1} involves some determinant,
which is an antisymmetric
function with respect to permutations of the integration variables
$z_1, \dots, z_s$. Because of
this antisymmetry, only the antisymmetric part of the double product,
with respect to permutations of these variables, actually contributes
to the multiple integral in \eqref{MIR1}.

Let us introduce the parametrization
\begin{equation}\label{zeds}
z_j=\omega(\xi_j),\qquad \tilde z_j = \tilde \omega(\xi_j)
\end{equation}
where the functions $\omega(\xi)=\omega(\xi;\lambda,\eta)$ and
$\tilde\omega(\xi)=\tilde \omega(\xi;\lambda,\eta)$
are as in \eqref{omegas} (note that upon this parametrization relation
\eqref{tildez} is satisfied automatically).
With this parametrization the double product in \eqref{MIR1}
coincides with the expression whose antisymmetric part has been
found in paper \cite{KMT-00}
(see appendix C of that paper). Namely, using a mixed set of notations,
the antisymmetric part of the double product, with respect to permutations of
the variables $z_1,\dots, z_s$, can be written as
\begin{multline}\label{asym}
\Asym_{z_1,\dots,z_s}
\prod_{1\leq j < k \leq s}^{}
\frac{(\tilde z_j-1)(z_k-1)}{\tilde z_j z_k -1}
\\
=\frac{1}{s!}\prod_{1\leq j < k \leq s}^{} (z_k-z_j)
\prod_{\substack{j,k=1 \\ j\ne k}}^{s}\frac{1}{b^2z_jz_k-(a^2+b^2-c^2)z_j+a^2}
\\ \times
\frac{a^{s(s-1)}c^{s(s-2)}}{\prod_{j=1}^{s} [\sin(\xi_j-2\eta)]^{s-1}}\,
Z_s(\lambda+\xi_1,\dots,\lambda+\xi_s).
\end{multline}
Here $Z_s(\lambda_1,\dots,\lambda_s)$ denotes the
partition function of the `partially' inhomogeneous
six-vertex model with domain wall boundary conditions on the $s\times s$
lattice; by `partially' inhomogeneous we mean that
the weights of this model are given by
\eqref{abc-im} and \eqref{abc}, but with all $\nu$'s equal zero.

The expression in the last line of \eqref{asym} is some symmetric polynomial
in variables $z_1,\dots,z_s$. Our aim below will be to show that this expression
admits a representation in terms of some determinant. Furthermore,
due to this representation it turns out possible to clarify the meaning
of the determinant in \eqref{MIR1} which appears to be related to the
partition function of the partially inhomogeneous model
on the $N\times N$ lattice with $s$ inhomogeneities (out of $N$ possible).

\subsection{Back to the partition function}

Let us consider the `partially' inhomogeneous six-vertex model
on $N\times N$ lattice, whose partition function will be denoted as
$Z_N(\lambda_1,\dots,\lambda_N)$.
The weights of the model are given by \eqref{abc-im}
and \eqref{abc}, with $\nu_1=\nu_2=\dots=\nu_N=0$, while
$\lambda_1,\lambda_2,\dots,\lambda_N$ are kept in general to be different.
Choosing some particular $\lambda$ as a value for
$\lambda_1,\lambda_2,\dots,\lambda_N$ in the homogenous limit,
one can introduce variables $\xi_1,\xi_2,\dots,\xi_N$ by
\begin{equation}\label{xi}
\lambda_\alpha=\lambda+\xi_\alpha,
\end{equation}
i.e., such that the homogenous limit will correspond
to making all $\xi$'s vanish. To simplify notations
we shall write $a(\lambda_\alpha):=a(\lambda_\alpha,0)$ and
$b(\lambda_\alpha):=b(\lambda_\alpha,0)$. Since both the homogeneous and
inhomogeneous models will be considered here simultaneously,
we shall write $a(\lambda)$ and $b(\lambda)$ for the
weights of the homogenous model (where $\lambda$ is
the same as in \eqref{xi}), rather than simply $a$ and $b$. We come back however
to these simplified notations in the last subsection.

An important role below will be played by
the variables $u_1,\dots,u_N$, defined by
\begin{equation}\label{ujs}
u_j=\gamma(\xi_j),\qquad
\gamma(\xi)=\frac{a(\lambda)}{b(\lambda)} \frac{b(\lambda+\xi)}{a(\lambda+\xi)}.
\end{equation}
For later use we mention here that function $\gamma(\xi)$
is related to function $\omega(\xi)$ \eqref{omegas}
via $\gamma(\xi)=\omega(-\lambda+\eta-\xi)$.

Our aim here will be to study the quantity $\wt Z_N(\xi_1,\dots,\xi_N)$, which
depends also on $\eta$ and $\lambda$ as parameters, and is defined by
\begin{equation}
Z_N(\lambda_1,\dots,\lambda_N)=Z_N(\lambda,\dots,\lambda) \wt Z_N(\xi_1,\dots,\xi_N).
\end{equation}
As we shall see, there is a nice representation
for this ``bare'' partition function in terms of variables \eqref{ujs}.

To illustrate the idea, it is useful to consider first the case when only
one inhomogeneity is present. Without lack of generality we can assume that this
is the inhomogeneity of the first row (i.e., $\xi_2=\cdots=\xi_N=0$).
The domain wall boundary conditions admit only one vertex of weight $c$
in the first row; if this vertex is at $r$-th position
(counted, as usual, from the right) then the first
$(r-1)$ vertices are of weight $b$ while the remaining
$(N-r)$ vertices are of weight $a$.
Taking this into account one has therefore the expression
\begin{equation}
Z_N(\lambda_1,\lambda,\dots,\lambda)=
\sum_{r=1}^{N}
\left[\frac{a(\lambda_1)}{a(\lambda)}\right]^{N-r}
\left[\frac{b(\lambda_1)}{b(\lambda)}\right]^{r-1}
H_N^{(r)}(\lambda,\dots,\lambda)Z_N(\lambda,\dots,\lambda).
\end{equation}
Note that here $H_N^{(r)}(\lambda,\dots,\lambda)$ denotes the
correlation function of the homogeneous model (recall that this function
describes the probability
of having the $c$-weight vertex at $r$-th position on the first row). Recalling
definition \eqref{hNz}, this last expression implies that
\begin{equation}\label{ZNone}
\wt Z_N(\xi_1,0,\dots,0)=
\left[\frac{a(\lambda_1)}{a(\lambda)}\right]^{N-1} h_N(u_1).
\end{equation}
It is to be emphasized that the variables $\xi_1$, $\lambda_1$, and $u_1$
are related to each other by \eqref{xi} and \eqref{ujs}; $\lambda$ is to be
regarded as a parameter, entering  also function
$\gamma(\xi)$ in \eqref{ujs}.

To consider the general case, let us introduce functions
$h_{N,s}(u_1,\dots,u_s)$, where the second subscript in the notation
refers to the number of arguments ($s=1,\dots,N$)
\begin{multline}\label{hNs}
h_{N,s}(u_1,\dots,u_s) =
\prod_{1\leq j<k \leq s}^{} (u_k-u_j)^{-1}
\\ \times
\begin{vmatrix}
u_1^{s-1} h_{N-s+1}(u_1)&  \hdots & u_s^{s-1} h_{N-s+1}(u_s) \\
u_1^{s-2}(u_1-1) h_{N-s+2}(u_1)& \hdots
&  u_s^{s-2}(u_s-1) h_{N-s+2}(u_s) \\
\hdotsfor{3} \\
(u_1-1)^{s-1} h_{N}(u_1)& \hdots
& (u_s-1)^{s-1} h_{N}(u_s)
\end{vmatrix}.
\end{multline}
These functions are symmetric polynomials of degree $(N-1)$
in each of their variables. They satisfy the relation (recall that $h_N(1)=1$)
\begin{equation}\label{reduction}
h_{N,s+1}(u_1,\dots,u_s,1)=h_{N,s}(u_1,\dots,u_s).
\end{equation}
This relation says that these functions can be in fact constructed
iteratively, starting with $h_{N,N}(u_1,\dots,u_N)$; one has then,
in particular,
$h_{N,1}(u_1)=h_{N}(u_1)$. The ``bare'' partition function,
in the most general case of $N$ inhomogeneities, reads
\begin{equation}\label{ZNxis}
\wt Z_N(\xi_1,\dots,\xi_N)=
\prod_{j=1}^{N} \left[\frac{a(\lambda_j)}{a(\lambda)}\right]^{N-1}
h_{N,N}(u_1,\dots,u_N).
\end{equation}
Using relation \eqref{reduction} one can easily obtain the corresponding formula
in the  case of $s$ (out of $N$) inhomogeneities.

It is worth to mention that
$h_{N,s}(u_1,\dots,u_s)$ introduced in \eqref{hNs}, besides
describing the partition function with $s$ inhomogeneities,
turns out to coincide,
modulo a trivial factor and the substitution $u_j\mapsto z_j$,
with the determinant in \eqref{MIR1}. We will exploit this observation
below in discussing equivalent multiple integral representations for EFP.

\subsection{Proof of the formula for $\wt Z_N(\xi_1,\dots,\xi_N)$}

Representation \eqref{ZNxis} for $\wt Z_N(\xi_1,\dots,\xi_N)$
(for the special case of $\lambda=\pi/2$)
appeared for the first time in paper \cite{CP-06}; since the proof
in that paper was lacking, we sketch it here for  completeness.

Let us begin with Izergin-Korepin formula, specialized to the case of
$\nu_1=\dots=\nu_N=0$, which reads
\begin{equation}
Z_N(\lambda_1,\dots,\lambda_N)=
\frac{\prod_{j=1}^N [a(\lambda_j)b(\lambda_j)]^N}{
\prod_{1\leq j<k\leq N}d(\lambda_k,\lambda_j)\prod_{n=0}^{N-1} n!}
\begin{vmatrix}
\varphi(\lambda_1)&  \hdots & \varphi(\lambda_N) \\
\partial_{\lambda_1}\varphi(\lambda_1)&  \hdots & \partial_{\lambda_N}
\varphi(\lambda_N) \\
\hdotsfor{3} \\
\partial_{\lambda_1}^{N-1}\varphi(\lambda_1)&  \hdots & \partial_{\lambda_N}^{N-1}
\varphi(\lambda_N)
\end{vmatrix}.
\end{equation}
Taking into account
that $\varphi(\lambda)=c/a(\lambda)b(\lambda)$, and
passing to the orthogonal
polynomial representation (see formulae \eqref{detdet} and \eqref{Knx}), we get
\begin{multline}
\wt Z_N(\xi_1,\dots,\xi_N)=
\frac{[\varphi(\lambda)]^{N(N-1)/2}}{
\prod_{j=1}^N [\varphi(\lambda_j)]^{N}
\prod_{1\leq j<k\leq N}d(\lambda_k,\lambda_j)}
\\ \times
\begin{vmatrix}
K_{0}(\partial_{\lambda_1}) &\hdots & K_{0}(\partial_{\lambda_N})\\
K_{1}(\partial_{\lambda_1}) &\hdots & K_{1}(\partial_{\lambda_N})\\
\hdotsfor{3} \\
K_{N-1}(\partial_{\lambda_1}) &\hdots & K_{N-1}(\partial_{\lambda_N})
\end{vmatrix}
\prod_{j=1}^{N}\varphi(\lambda_j).
\end{multline}
Now, taking into account that
\begin{equation}\label{uud}
u_k-u_j =\frac{[a(\lambda)]^2}{a(\lambda_j)a(\lambda_k)}\,
\varphi(\lambda) d(\lambda_k,\lambda_j)
\end{equation}
and
\begin{equation}\label{aapp}
\frac{\varphi(\lambda) a(\lambda)}{\varphi(\lambda_j) a(\lambda_j)}
=\frac{a(\lambda_j)}{a(\lambda)}\, u_j,
\end{equation}
we arrive at
\begin{multline}\label{almostdone}
\wt Z_N(\xi_1,\dots,\xi_N)=
\prod_{j=1}^{N} \left[\frac{a(\lambda_j)}{a(\lambda)}\right]^{N-1}
\frac{\prod_{j=1}^{N}u_j^{N-1}}{\prod_{1\leq j<k\leq N}(u_k-u_j)}
\\ \times
\frac{1}{\prod_{j=1}^N \varphi(\lambda_j)}
\begin{vmatrix}
K_{0}(\partial_{\lambda_1}) &\hdots & K_{0}(\partial_{\lambda_N})\\
K_{1}(\partial_{\lambda_1}) &\hdots & K_{1}(\partial_{\lambda_N})\\
\hdotsfor{3} \\
K_{N-1}(\partial_{\lambda_1}) &\hdots & K_{N-1}(\partial_{\lambda_N})
\end{vmatrix}
\prod_{j=1}^{N}\varphi(\lambda_j).
\end{multline}
Comparing with \eqref{ZNxis}, we see that to
complete the proof we only need to express the second line in terms of $u$'s.

For this purpose let us again turn to the special case when only one
inhomogeneity is present. In this case Izergin-Korepin formula boils down to
\begin{multline}
Z_N(\lambda_1,\lambda,\dots,\lambda)=
\frac{[a(\lambda_1)b(\lambda_1)]^N[a(\lambda)b(\lambda)]^{N(N-1)}}{
(N-1)!\prod_{n=0}^{N-2} (n!)^2 [d(\lambda_1,\lambda)]^{N-1}}
\\ \times
\begin{vmatrix}
\varphi(\lambda)& \dots & \partial_\lambda^{N-2}\varphi(\lambda) & \varphi(\lambda_1)
\\
\partial_\lambda\varphi(\lambda) & \dots & \partial_\lambda^{N-s}\varphi(\lambda)
 & \partial_{\lambda_1} \varphi(\lambda_1)
\\ \hdotsfor{4} \\
\partial_\lambda^{N-1}\varphi(\lambda) & \dots &
\partial_\lambda^{2N-3}\varphi(\lambda) &
\partial_{\lambda_1}^{N-1} \varphi(\lambda_1)
\end{vmatrix}.
\end{multline}
We have
\begin{equation}
\wt Z_N(\xi_1,0,\dots,0)=
\frac{1}{[\varphi(\lambda_1)]^N[d(\lambda_1,\lambda)]^{N-1}}\,
K_{N-1}(\partial_{\lambda_1})\varphi(\lambda_1).
\end{equation}
Taking into account that \eqref{uud} implies that
$d(\lambda_1,\lambda)=(u_1-1)a(\lambda_1)[a(\lambda)\varphi(\lambda)]^{-1}$
and also using \eqref{aapp}, we have
\begin{equation}
\wt Z_N(\xi_1,0,\dots,0)=
\left[\frac{a(\lambda_1)}{a(\lambda)}\right]^{N-1}
\left(\frac{u_1}{u_1-1}\right)^{N-1} \frac{1}{\varphi(\lambda_1)}\,
K_{N-1}(\partial_{\lambda_1}) \varphi(\lambda_1).
\end{equation}
Now, comparing this expression with \eqref{ZNone}, we find that
\begin{equation}
\frac{1}{\varphi(\lambda_j)}\, K_n(\partial_{\lambda_j})\varphi(\lambda_j)=
\bigg(\frac{u_j-1}{u_j}\bigg)^n h_{n+1}(u_j).
\end{equation}
Finally, plugging this formula into \eqref{almostdone}, one arrives at \eqref{ZNxis},
which is thus proven.

\subsection{Equivalent representations}

Coming back to the expression in the last line of formula \eqref{asym},
one immediately finds, using \eqref{ZNxis}
(and also taking into account \eqref{zeds},
\eqref{ujs}, and \eqref{go}) that
\begin{equation}\label{last}
\frac{Z_s(\lambda+\xi_1,\dots,\lambda+\xi_s)}
{\prod_{j=1}^{s} [\sin(\xi_j-2\eta)]^{s-1}}
= \frac{Z_s }{c^{s(s-1)}} \prod_{j=1}^{s} \left(\frac{z_j-1}{u_j}\right)^{s-1}
h_{s,s}(u_1,\dots,u_s).
\end{equation}
Here $Z_s$ stands for $Z_s(\lambda,\dots,\lambda)$ and below we also come back to
our simplified notations ($a=a(\lambda,0)$, $b=b(\lambda,0)$). Furthermore, we shall
use the following notations
\begin{equation}
\Delta:=\frac{a^2+b^2-c^2}{2ab}=\cos(2\eta),\qquad
t:=\frac{b}{a}=\frac{\sin(\lambda-\eta)}{\sin(\lambda+\eta)}.
\end{equation}
Exploiting formulae \eqref{omegas} and \eqref{ujs}
it is straightforward to find that
\begin{equation}
u_j=-\frac{z_j-1}{(t^2-2t\Delta)z_j+1}.
\end{equation}
This formula, by the way,
shows that \eqref{last} is indeed a polynomial in $z_1,\dots,z_s$ (recall that
in general the functions $h_{N,s}(u_1,\dots,u_s)$ are symmetric polynomials
of degree $(N-1)$ in each variable). In all,
for the antisymmetrical part of the double product in \eqref{MIR1}, we obtain
\begin{multline}\label{asymtot}
\Asym_{z_1,\dots,z_s}
\prod_{1\leq j < k \leq s}^{}
\frac{(\tilde z_j-1)(z_k-1)}{\tilde z_j z_k -1}
=\frac{Z_s}{s!a^{s(s-1)} c^s}\prod_{1\leq j < k \leq s}^{} (z_k-z_j)
\\ \times
\frac{\prod_{j=1}^{s} [(t^2-2t\Delta)z_j+1]^{s-1}\,
h_{s,s}(u_1,\dots,u_s)}{\prod_{\substack{j,k=1 \\ j\ne k}}^{s}
(t^2z_jz_k-2t\Delta z_j+1)}.
\end{multline}
Taking account that the
determinant in \eqref{MIR1} can be represented in terms of the
function $h_{N,s}(z_1,\dots,z_s)$, we arrive therefore at the
following equivalent multiple integral representation for EFP
\begin{multline}\label{MIR2}
F_N^{(r,s)}=\frac{(-1)^{s(s+1)/2}Z_s}{s!(2\pi\rmi)^s a^{s(s-1)}c^s}
\oint_{C_0}^{} \cdots \oint_{C_0}^{} \prod_{1\leq j<k\leq s}^{} (z_k-z_j)^2
\\ \times
\prod_{\substack{j,k=1\\ j\ne k}}^{s} \frac{1}{t^2 z_jz_k-2t\Delta z_j +1}
\prod_{j=1}^{s} \frac{[(t^2-2t\Delta)z_j+1]^{s-1}}{z_j^r(z_j-1)^s}
\\ \times
h_{N,s}(z_1,\dots,z_s)h_{s,s}(u_1,\dots,u_s)
\,\rmd z_1\cdots \rmd z_s.
\end{multline}
This formula is the main result of the present paper.

As a comment to this result, let us also mention that due to
\eqref{ZNxis} one can obtain from \eqref{MIR2} a
multiple integral representation, in which the integrand is expressed in terms
of the partition functions. Indeed, since $z_j$'s
and $u_j$'s are given by \eqref{omegas} and \eqref{ujs} where
functions $\omega(\xi)$ and $\gamma(\xi)$ appear to be related to each other
as $\gamma(\xi)=\omega(-\lambda+\eta-\xi)$, or
$\omega(\xi)=\gamma(-\lambda+\eta-\xi)$, one may prefer to use $\xi_1,\dots,\xi_s$
as the integration variables, and furthermore \eqref{ZNxis} implies that
\begin{equation}
h_{N,s}(\omega(\xi_1),\dots,\omega(\xi_s))=
\frac{Z_N(\eta-\xi_1,\dots,\eta-\xi_s,\lambda,\dots,\lambda)}
{Z_N(\lambda,\dots,\lambda)}
\prod_{j=1}^{s} \left[\frac{a(\lambda)}{a(\eta-\xi_j)}\right]^{N-1}.
\end{equation}
As a result, we obtain the following alternative multiple integral representation
\begin{multline}\label{MIR3}
F_N^{(r,s)}=\frac{(-1)^{Ns+s(s+1)/2}a^{(N-r)s}b^{rs}}{s!(2\pi\rmi)^s c^sZ_N}
\oint_{C_0}^{} \cdots \oint_{C_0}^{} \prod_{1\leq j<k\leq s}^{} [\sin(\xi_k-\xi_j)]^2
\\ \times
\prod_{\substack{j,k=1\\ j\ne k}}^{s} \frac{1}{\sin(\xi_j-\xi_k+2\eta)}
\prod_{j=1}^{s} \frac{1}{[\sin(\xi_j-2\eta)]^{N-r}
[\sin(\xi_j+\lambda-\eta)]^s(\sin\xi_j)^r}
\\ \times
Z_N(\eta-\xi_1,\dots,\eta-\xi_s,\lambda,\dots,\lambda)
Z_s(\lambda+\xi_1,\dots,\lambda+\xi_s)
\,\rmd \xi_1\cdots \rmd \xi_s.
\end{multline}
This formula evidently recalls multiple integral
representations which appear for correlation functions in
quantum spin chains (see, e.g., \cites{JM-95,KMT-00,BKS-03,BJMST-06}).

The important and essentially nontrivial object in the integrand of \eqref{MIR3}
is the partition function $Z_N$, with $s$ shifted arguments. While the presence
of the second one, $Z_s$, is a standard consequence of the
Yang-Baxter algebra, the first partition function, $Z_N$,
together with
other factors in the integrand, is due to the specificities of
both the boundary conditions (domain-wall ones)
and the particular correlation function under consideration (EFP).

In conclusion, let us briefly discuss
formula \eqref{MIR2}. This representation may seem not completely explicit
since it is heavily based on the use of the generating function $h_N(z)$.
We expect however representation \eqref{MIR2} to appear most useful
when addressing the problem of limit shapes in the model.
In the case of  vertex weights obeying the free-fermion condition ($\eta=\pi/4$),
function $h_N(z)$ is known explicitly and particularly simple.
This case has been considered in \cite{CP-07}. Fortunately,
function $h_N(z)$  is also known at  $\eta=\pi/6$ and $\eta=\pi/3$,
for $\lambda=\pi/2$. The first case corresponds
to enumeration of alternating sign matrices and the
second one to a particular example of their weighted enumeration
(the so-called $3$-enumeration). For this reason we expect that
formula \eqref{MIR2} may appear also very useful out of the free-fermion case,
and, at least, allows one to solve the long-standing problem
of the limit shape of large alternating sign matrices.

\section*{Acknowledgements}

One of us (AGP) is supported in part by Russian Foundation for Basic
Research, under grant 07-01-00358, and by the program ``Mathematical
Methods in Nonlinear Dynamics'' of Presidium of Russian Academy of
Sciences. AGP also thanks INFN, Sezione di Firenze, where part of this work
was done, for hospitality and support. This work is
partially done within the European Science Foundation program INSTANS.

\bibliographystyle{amsplain}
\bibliography{efp_bib}

\end{document}